\documentclass[conference]{IEEEtran}
\IEEEoverridecommandlockouts

\usepackage{cite}
\usepackage{amsmath,amssymb,amsfonts,amsthm}
\usepackage{algorithm2e}
\usepackage{graphicx}
\usepackage{textcomp}
\usepackage{xcolor}
\def\BibTeX{{\rm B\kern-.05em{\sc i\kern-.025em b}\kern-.08em
    T\kern-.1667em\lower.7ex\hbox{E}\kern-.125emX}}

\DeclareMathOperator*{\argmax}{arg\,max}

\newtheorem{definition}{Definition}

\newtheorem{theorem}{Theorem}
\begin{document}

\title{Probabilistic Time Series Forecasting     of Residential Loads -- A Copula Approach}

\author{\IEEEauthorblockN{1\textsuperscript{st} Marco Jeschke}
\IEEEauthorblockA{\textit{Department of Statistics} \\
\textit{TU Dortmund University}\\
Dortmund, Germany \\
marco.jeschke@tu-dortmund.de}
\and
\IEEEauthorblockN{2\textsuperscript{nd} Timm Faulwasser}
\IEEEauthorblockA{\textit{Institute of Control Systems} \\
\textit{Hamburg University of Technology}\\
Hamburg, Germany \\
timm.faulwasser@ieee.org}
\and
\IEEEauthorblockN{3\textsuperscript{rd} Roland Fried}
\IEEEauthorblockA{\textit{
Department of Statistics} \\
\textit{TU Dortmund University}\\
Dortmund, Germany\\
fried@statistik.tu-dortmund.de}

}

\maketitle\footnotetext{The authors acknowledge funding in the course of TRR 391 Spatio-temporal \textit{Statistics
for the Transition of Energy and Transport} (520388526) by the Deutsche
Forschungsgemeinschaft (DFG, German Research Foundation). }

\begin{abstract}
Predicting the time series of future evolutions of renewable injections and demands is of utmost importance for the operation of power systems. However, the current state of the art is mostly focused on mean-value time series predictions and only very few methods provide probabilistic forecasts. In this paper, we rely on kernel density estimation and vine copulas to construct probabilistic models for individual load profiles of private households. Our approach allows the  quantification of variability of individual energy consumption in general and of daily peak loads in particular. We draw upon an Australian
distribution grid dataset to illustrate our findings. We generate synthetic loads that follow the distribution of the real data.

\end{abstract}

\begin{IEEEkeywords}
load forecasting, probabilistic modeling, distribution grids, kernel density estimation, vine copulas
\end{IEEEkeywords}

\section{Introduction}
Due to increasing amounts of renewables and growing electricity demand, ensuring the stability of electricity supply in power grids is becoming an increasingly important yet complex task for grid operators. To guarantee this stability of supply, adequate models and reliable predictions of  electricity consumption patterns are needed. Volatile renewable energy systems and uncertainties in consumption behavior pose challenges for energy modeling and forecasting. Yet, it is crucial to manage and quantify the uncertainty surrounding  forecasts.
Moreover, the ongoing trend towards increasing automation in distribution systems and active coordination of distribution and transmission systems induces the need for time series forecasting on lower voltage levels, see, e.g., \cite{forecasting}.

Forecasting methods can be grouped into point forecasts and probabilistic forecasts. The former predict the expected value but do not provide information about the uncertainty surrounding the forecast.  Hence, probabilistic forecasts are gaining importance \cite{prohabilistic}. Common techniques for probabilistic forecasting of loads, wind energy and photovoltaics include kernel-based methods \cite{kernelSVR,kernelNonPara}, quantile regression methods \cite{SplineQ,SplineSVR,QuantileReg,QuantileReg2,QuantileReg3}, neural networks \cite{PVNN}, support vector regression \cite{kernelSVR,SplineSVR}, and combinations of these. 
An alternative avenue  is the integration of copulas into the aforementioned techniques. Copulas are mathematical functions used to describe and model the dependency structure between random variables. They separate the marginal distributions of individual variables from their joint dependency, allowing for flexible modeling of complex relationships. By linking univariate marginal distributions to form a multivariate distribution, copulas enable the study of dependencies beyond simple linear correlations, making them particularly useful in fields like finance, hydrology, and energy modeling. 
 This allows dependency structures to be incorporated into the various forecasting models \cite{Copula1,Copula2}. Copulas for modeling temporal dependency structures, that is, dependencies in load profiles at different times of the day, have been less considered in current works. 

An alternative to time series forecasting for modeling consumption are standard load profiles (SLP) \cite{SLP}, in addition to weather forecasts and other data. These profiles represent typical consumption behaviors of different types of end users. For Germany these SLPs were created by the German Federal Association of Energy and Water Management (Bundesverband der Energie- und Wasserwirtschaft e. V.) \cite{SLP2}. Other approaches model loads based on individual behavior \cite{Generator,Generator2}.

The quality of predictions based on such profiles depends heavily on how accurately the respective end user is represented by the load profile. 
However, the consumption behavior of  households depends on the habits of their residents \cite{SLPProblem}. Nowadays, many people can work  in home office, others work in shifts, and some own electric vehicles or photovoltaic (PV) systems which feed energy back into the grid. Such factors may influence consumption behavior substantially, causing actual power demands to deviate widely from  standard load profiles. Additionally, a standard load profile only represents average behavior and does not reflect its variability. Rare occurrences of extreme patterns are not accounted for.

This paper investigates modeling of individual electricity consumption curves that  accounts for the varying behavior of individual end users and considers extreme consumption patterns with a copula-based approach. The underlying motivation is the continued roll-out of smart meters.
The contribution of this paper is to incorporate temporal dependencies in consumption behavior into the modeling of electricity load curves.  
In contrast to previous works, we use a combined approach of kernel density estimation and vine copulas. On the one hand, this allows to identify density-based similarities in consumption behavior. Moreover, it relies on data-driven models of dependency structures for modeling. This has the advantage that we can account for high-dimensional dependencies in load profiles and we can provide probabilistic forecasts in the form of quantiles of the kernel density estimators for the resulting load profiles.

The remainder of the paper is structured as follows: We briefly describe the data and statistical methods used in this study in Section 2.
In Section 3, we generate individual load profiles capable of representing statistically extreme patterns. In Section 4, we give a summary of our results and an outlook on future work and open problems.

\section{Problem statement and preliminaries}
 We propose a computationally tractable approach for forecasting the future time series of household consumption  which also provides information about the probability of their occurrence. We focus on this microscopic problem  as a challenging benchmark with non-Gaussian statistics. To this end, we combine kernel- and quantile-based methods with a copula-based approach to model dependencies between consumption behavior at different point in time.

Specifically, we want to forecast active power  profiles of households. For each time point $t$, we aim to estimate a density $f_t$ that describes the distribution of active power $P(t)$. Using kernel density estimation, we describe $P(t)$ as a random variable with the density function $f_t$, i.e., $P(t)\sim f_t$. Because the active power at different times $t_1\neq t_2$, i.e., $P(t_1)$ and $P(t_2)$, may depend statistically on each other, we use Vine Copulas to model those dependencies.
\subsubsection{Data}
The dataset analysed in the following consists of electricity data collected by the Australian utility company Ausgrid \cite{b1}. These data were recorded with a resolution of 30 minutes over the period from July 1, 2010, to June 31, 2013, and include electricity consumption, photovoltaic (PV) generation, as well as data on electric hot water systems from a total of 300 private households. For the generation of load profiles we use only the consumption data. Each entry represents the electricity consumption of the
customer over the preceding half hour. The households are located in or around the city of Sydney in the state of New South Wales and are supplied by the same distribution network.

The 300  households were anonymized by Ausgrid and randomly selected from 10,000 previously collected customer datasets. Here, we only consider a subset of these 300 households. These 53 households were selected as a corrected dataset based on the approach in \cite{b1}. This ensures that the data are not distorted, e.g., by extended vacation periods of residents. For a detailed description of the selection criteria, we refer to \cite{b1}.

Henceforth, we focus on household \#1. Extending the approach to other households and generating additional individual load profiles is straightforward. Conceptually, our approach also readily extends to reactive power, yet the considered dataset does not provide such training data~\cite{b1}.

\subsection{Statistical methods} \label{sub:stat} 
We aim to identify statistical similarities as well as differences in consumption behavior. We estimate the underlying probability densities for modeling not only average consumption patterns but their full distribution. Moreover, we aim to identify time periods that exhibit statistical similarities. Such periods might include e.g. nighttime rest, shared wake-up times, or working hours. To identify such periods, the available data are divided into possibly homogeneous time blocks. These blocks are analyzed separately and then they are subsequently either aggregated or treated as distinct segments.

\subsubsection{Kernel density estimation}  
In order to avoid restrictive parametric distributional assumptions we estimate the probability density of the electricity consumption in a given time block using kernel density estimation (KDE). For this purpose, we recall  the following concepts from statistics. 

\begin{definition}[Kernel function] \label{def:kernel}
    A kernel is a non-negative real-valued function $K:\mathbb U \to \mathbb R$ with the following characteristics:
        \begin{align}\label{eq:kernel}
        \int_{-\infty}^{\infty}K(u)du&=1, \qquad
        K(-u)=K(u)~\forall u\in U.\
    \end{align}
\end{definition}
\noindent Examples  are
uniform kernels $K(u)\equiv\frac{1}{2}$ with support $[-1, 1]$ 

or  Gaussian kernels $K(u)=\frac{1}{\sqrt{2\pi}}\exp\lbrace-\frac{1}{2}u^{2}\rbrace$ with support $\mathbb{R}$.

\begin{definition}[Kernel density estimator  \cite{b2}]
Let $x_{1},...,x_{n}$ be independent and identically distributed samples from a univariate distribution with probability density $f(x)$. The kernel density estimator of $f$ is given by
\begin{equation}
    \hat{f}_{h}(x)=\frac{1}{nh}\sum_{i=1}^{n}K\biggl(\frac{x-x_{i}}{h}\biggr),\label{eq:kern}
\end{equation}
whereby the tuning parameter $h>0$ is called bandwidth.
\end{definition}
The choice of the kernel usually has little  impact on the estimator if the  smoothing parameter $h$ is chosen adequately  \cite{b2}. A value of $h$ that is very small may result in the estimator capturing irrelevant details, whereas a value of $h$ that is very large can lead to over-smoothing. Therefore, selecting an appropriate bandwidth is  crucial. Popular methods for this are Scott's rule $h\approx 1.06\hat{\sigma}n^{-\frac{1}{5}}$ or Silverman's rule  $h=0.9\min\bigl(\hat{\sigma},\frac{IQR}{1.35}\bigr)n^{-\frac{1}{5}}$, where $\hat{\sigma}$ is the sample standard deviation and $IQR$ is its interquartile range. Both methods work best with normally distributed samples.
Since the underlying distribution is unknown, we use the non-parametric plug-in estimator by Sheather and Jones, which is based on minimizing an approximation of the asymptotic mean integrated squared error \cite{b5}.

\subsubsection{Clustering}
 We want to identify similarities in consumption patterns and thus compare the  density estimates of households for commonalities. To this end, we consider the squared Hellinger distance as a statistical distance measure. 
 
\begin{definition}[Squared Hellinger distance \cite{Vaart}] Let $\mu$ and $\nu$ be two probability measures on a measurable space $\mathcal X$ that are absolutely continuous with respect to a measure $\lambda$. The squared Hellinger distance between $\mu$ and $\nu$ is defined as
\begin{equation}
   H^{2}(\mu,\nu)=\frac{1}{2}\int_{X}\biggl(\sqrt{f(x)}-\sqrt{g(x)}\biggr)^{2}\lambda(dx)=D_H(f,g),\label{eq:hell} 
\end{equation}
where $f$ and $g$ are the Radon-Nikodym derivatives of $\mu$ and $\nu$, respectively.
\end{definition}

The squared Hellinger distance satisfies all the properties of a distance measure, as required for clustering algorithms. We choose a k-means algorithm based on the squared Hellinger distance for the clustering. The quality of an assignment is determined using the silhouette coefficient.

\begin{definition}[Silhouette coefficient\cite{Sil}] \label{def:sil} Assume we have a division of the densities into $K$ clusters $C_{1},...,C_{K}$. The silhouette coefficient of a density $f$ from cluster $C_{I}$ is defined as 
\begin{equation}
    s(f)=\begin{cases} 1-\frac{a(f)}{b(f)},~a(f)\leq b(f) \\ \frac{b(f)}{a(f)}-1,~a(f)>b(f)\end{cases}.\label{eq:sil}
\end{equation}
Here, $a(f)=\frac{1}{\mid C_{I}\mid -1}\sum_{g\in C_{I},~g\neq f}D_{H}(f,g)$ is the average squared Hellinger distance between $P$ from cluster $C_{I}$ and all other densities from $C_{I}$, while $b(f)=\min_{J\neq I}\frac{1}{\mid C_{J}\mid}\sum_{g\in C_{J}}D_{H}(f,g)$ is the smallest average distance between $f$ and all densities from another cluster.
\end{definition}
A low value of $a$ indicates a good assignment, as does a high value of $b$. According to the definition, the silhouette coefficient can take a value in the interval $[-1,1]$. The larger the value, i.e., the smaller $a$ or the larger $b$, the better the assignment of the density $f$ to the cluster $C_{I}$.

The silhouette coefficient of a cluster $S(C)$ is defined as the average silhouette coefficient of all densities in this cluster. Generally, a value of $1 \geq S(C) > 0.75$ indicates a very homogeneous cluster, a value of $0.75 \geq S(C) > 0.5$ indicates  good to moderate homogeneity, and a value  $S(C) \leq 0.5$ indicates a poor structure. For each number of clusters $K$ considered, we examine the average coefficient across all clusters $\overline{S_{K}(C)}$. A suitable number of clusters can then be determined as $\argmax_{K}\overline{S_{K}(C)}$.

\subsubsection{Vine Copulas}\label{VineCop}  
Our underlying assumptions is that there are certain dependencies in the consumption patterns within a single day. For example, if a person  cooks dinner at 6:00 PM, this will likely affect the energy consumption at 6:30 PM and possibly even at 7:00 PM. In the case of a dishwasher or washing machine, even longer lasting dependencies over time can emerge. To account for such time dependencies in the load profiles, we make use of so-called Vine Copulas.

An $n$-dimensional copula is a multivariate distribution function $C:[0,1]^{n}\rightarrow [0,1]$ with uniformly distributed marginal distributions. It allows a detailed analysis of stochastic dependencies by separating the marginals from the dependencies. The foundation is Sklar's  theorem.  

\begin{theorem}[Sklar\cite{b4}] Let $F$ be an $n$-dimensional distribution function with marginal distributions $F_{1},...,F_{n}$. Then, there exists an $n$-dimensional copula 
$C:[0,1]^{n}\rightarrow [0,1]$ such that  
$$F(x_{1},...,x_{n})=C(F_{1}(x_{1}),...,F_{n}(x_{n})).$$  If the multivariate distribution has a density $f$ we have 
\begin{equation}
    f(x_{1},...,x_{n})=c(F_{1}(x_{1}),...,F_{n}(x_{n}))f_{1}(x_{1})\cdots f_{n}(x_{n}),\label{eq:sklar}
\end{equation}
for all $x_{i}\in\mathbb{R}\cup\{-\infty,\infty\}$ with the copula density $c$. 
\end{theorem}  
For absolutely continuous distributions, the copula $C$ is even unique. There is a wide variety of bivariate copulas available for modeling different types of dependencies. 

Because copula densities are usually unbounded at the edges of $[0,1]^{2}$ we restrict the copula densities at the edges of $[0,1]^{2}$ through a normalization of the marginals for a better comparison.

We consider the family of so-called Vine Copulas where multivariate density functions are decomposed into conditional densities and marginal distributions. The conditional densities can then be represented by bivariate copulas. \cite{b3}

\begin{figure}[t] 
\centering
 \includegraphics[width=0.4\textwidth,height = 2cm]{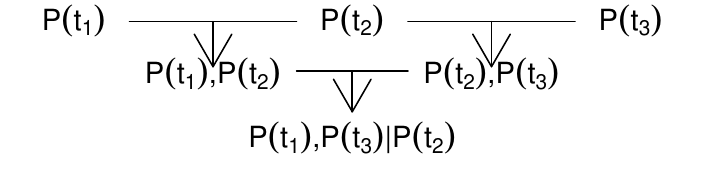} 
\caption{\label{Fig:VineTree}3-dimensional vine structure for active power}
\end{figure}

In the two-dimensional case, we have 
\begin{equation}
   f(x_1,x_2)=f_{1\mid 2}(x_{1}\mid x_{2})f_{2}(x_{2})=f_{2\mid 1}(x_{2}\mid x_{1})f_{1}(x_{1})\label{eq}
\end{equation}
 with $f_{1\mid 2}(x_{1}\mid x_{2})=c_{1,2}(F_{1}(x_{1}),F_{2}(x_{2}))f_{1}(x_{1})$ and $f_{2\mid 1}(x_{2}\mid x_{1})=c_{1,2}(F_{1}(x_{1}),F_{2}(x_{2}))f_{2}(x_{2})$.\\ 
This decomposition forms the basis of Vine Copulas. It becomes apparent already in the two-dimensional case  that this decomposition is not unique. For $d=3$ there are already $6$  possible decompositions. Using the recursive representations $f(x_1,x_2,x_3)=f_{1\mid 2,3}(x_{1}\mid x_{2},x_{3})f_{2, 3}(x_{2}, x_{3})$ and $f_{2, 3}(x_{2}, x_{3})=f_{2\mid 3}(x_{2}\mid x_{3})f_{3}(x_3)$ one of these possibilities is $f(x_1,x_2,x_3)=f_{1\mid 2,3}(x_{1}\mid x_{2},x_{3})f_{2\mid 3}(x_{2}\mid x_{3})f_{3}(x_3)$. Applying Sklar's theorem gives 
\begin{align} 
&f(x_1,x_2,x_3)=f_{1}(x_{1})f_{2}(x_{2})f_{3}(x_{3}) \notag \\
&~~~~~~~~~~~~~~~~~~~~\times c_{2,3}(F_{2}(x_{2}),F_{3}(x_{3}))c_{1,3}(F_{1}(x_{1}),F_{3}(x_{3})) \notag \\ 
&~~~~~~~~~~~~~~~~~~~~\times c_{1,2\mid 3}(F_{1\mid 3}(x_{1}\mid x_{3}),F_{2\mid 3}(x_{2}\mid x_{3})).\label{eq}
\end{align} 
This decomposition reflects the tree structure shown in Figure \ref{Fig:VineTree}. First, the dependencies between variables $1$ and $3$ and between $2$ and $3$ are modeled using bivariate copulas. Then, the conditional distribution functions that result from this are again connected using a bivariate copula. It is known that for $n$ variables, there are $\frac{n!}{2}2^{\binom{n-2}{2}}$ possible decompositions. Later, we will see that in our case, the tree structure arises naturally.

We can think of this in terms of the active power on three consecutively time points $t_1,t_2,t_3$. The value of $P(t_2)$ depends on the value of $P(t_1)$ and the value of $P(t_3)$ depends on $P(t_2)$ and maybe even on $P(t_1)$. With the help of the marginal densities $f_{t_1},f_{t_2},f_{t_3}$ we can model the dependencies pairwise with a three dimensional Vine.

The statistical methods presented above result in a computationally tractable procedure that provides individual load profiles for a household, see Algorithm~\ref{algo:KDE}.

\RestyleAlgo{ruled}
\begin{algorithm}
    \caption{KDE for individual household loads}
    \label{algo:KDE}

    \KwData{$N$ load data points with length $T$}
    \For{$t\in 1:T$}{Estimate the densities of the consumption at time point $t$ with \eqref{eq:kern}\;}
     Cluster the densities to divide the day into $K$ states with \eqref{eq:hell} and \eqref{def:sil}\;
     \For{$k\in 1:K$}{Determine the dependency structure for state $k$ with \eqref{eq:sklar}\;}
    Generate data that follow these structures and transform them with help of the estimated densities and their quantile functions, respectively\;
    \textbf{Output}: Any number of simulated load profiles for a certain household\;

\end{algorithm}

\subsection{Permutation test}\label{VineCop}  
To validate our generated load profiles, we apply a permutation test. This allows us to compare the simulated loads with the real data. For this purpose, we compare the distance between various statistical characteristics of the real data and the simulated data.

\begin{definition}[Permutation test]
 Let $(q_{i}^{1},...,q_{i}^{n})=x_{i}$ with $1\leq i\leq N$ be a vector of $n$ statistical features $q^1,\ldots,q^n$ for $N$ real observations and let $(q_{i}^{1},...,q_{i}^{n})=x_{i}$ with $N+1\leq i\leq 2N$ be a vector of the same features for $N$ simulated observations. We define the distance between two  statistical feature vectors $x_{i}$ and $x_{j}$ as $d(i,j)=(x_{i}-x_{j})^\top\Sigma^{-1}(x_{i}-x_{j})$ with $\Sigma=\frac{1}{2}(\Sigma_{1}+\Sigma_{2})$ and $\Sigma_{i}$ the covariance matrix of the feature vectors for the real respectively simulated observations. The resulting test statistic is defined as 
 \[T=\sum_{i\in I_{1}, j\in I_{2}}d(i,j) \quad \text{with} \quad I_{1}=\{1,...,N\}
 \]
 and $I_{2}=\{N+1,...,2N\}$. We then compute $M$ permuted test statistics with random partitions  $\{1,...,2N\}=I_1^j\cup I_2^j$ into subsets $I_1^j$ and $I_2^j$ with $\mid I_{1}^j\mid=N=\mid I_{2}^j\mid$. The p-value is the proportion of permuted test statistics that are larger than the original test statistic.
 \end{definition}

\section{Case study on Ausgrid Data}

Due to space limitations, we only consider the Australian winter months (June, July, August) and working days (Monday to Thursday). We take possible daily  seasonal variation of electricity consumption into account, estimating  the probability density for each time point on a given day separately. Since we have measurements every $30$ minutes, this corresponds to  $48$ possibly different densities during each day. For kernel density estimation, we need to ensure that we have enough data points available. To achieve this, we assume in a first step that the data follow the same daily pattern from Monday to Thursday during the winter months. This assumption leads to N=$156$ data samples for each of these time points, see Table \ref{tab2}. Friday, Saturday and Sunday and the other months need to be considered separately because of a different consumption behavior, using an analogous approach. 

\begin{table}[h!]
 \caption{Number of data points used for kernel density estimation}
     \centering
    \begin{tabular}{c||c|c|c||c} 
        \textbf{~} & \textbf{June} & \textbf{July} & \textbf{August} & \textbf{~} \\
        \hline
        Monday & 12 & 13 & 14 &  39\\
        \hline
        Tuesday & 12 & 13 & 14 &  39\\
        \hline
        Wednesday & 13 & 12 & 14 &  39\\
        \hline
        Thursday & 13 & 13 & 13 &  39\\
        \hline
        \hline 
         & 50 & 51 & 55 & 156 \\
    \end{tabular}
    \label{tab2}
    \end{table}

 We use a Gaussian kernel and the bandwidth estimator by Sheather and Jones. Since we expect to observe more values smaller than the mean for active power, we can assume a right-skewed distribution of the data. Because kernel density estimation works better with symmetric data, we first apply a logarithmic transformation to the data.
 
\begin{figure}[t]
\centering
    \includegraphics[width=0.5\textwidth, height = 6cm]{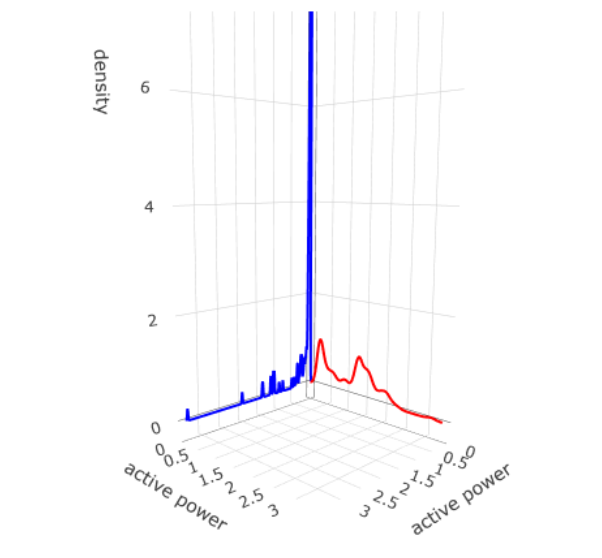} 
    \caption{\label{Fig:Density}Two examples for (non-normalized) densities of different clusters. The blue density as an example for cluster 1 and the red density for cluster 2.}
\end{figure}

Figure \ref{Fig:Density} presents an example of two densities (for 1 a.m. and 5 p.m.) of different clusters. We can see that the active powers follow different distributions. This was to be expected, as the consumption at 1 a.m. certainly differs substantially from the consumption at 5 p.m. Note the non-Gaussianity for both densities. The density from Cluster 1 exhibits a strong right skew, while the density from Cluster 2 shows two peaks. That is, one may use a Gaussian kernel to estimate non-Gaussian densities.

In the next step, we examine which of these densities resemble each other and whether we can find certain groups. This allow us to estimate the density for certain time periods of the day jointly using more data points, resulting in less volatile estimates.  

\begin{figure}[t!]
  {\includegraphics[width=0.49\textwidth,height = 4cm]{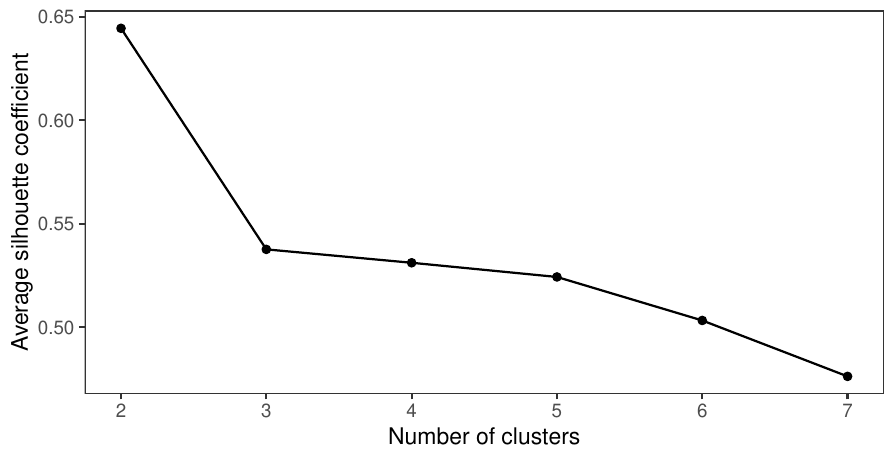}}
  \caption{\label{Fig:cluster} Average silhouette coefficient for different numbers of clusters. The optimal choice is two clusters with a coefficient of 0.61 indicating high homogeneity within the clusters. }
\end{figure}

\begin{figure}[t]
  
  {\includegraphics[width=0.49\textwidth,height = 4cm]{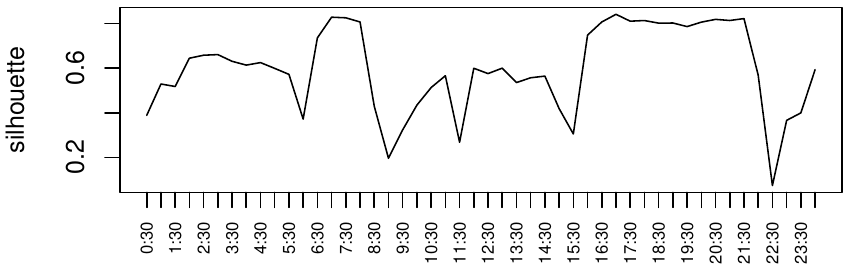}}
\caption{\label{Fig:sil}Silhouette for every density}
\end{figure}

Figure \ref{Fig:cluster} indicates that the optimal number of clusters is $2$ with an average silhouette coefficient of $0.61$, which corresponds to a homogeneous clustering. Table \ref{tab3} shows the resulting cluster assignment. The periods 22:30-6:00 and 8:30-15:30 are assigned to the same cluster, just like 6:30-8:00 and 16:00-22:00. In Figure \ref{Fig:sil}, we see that the silhouette coefficient drops sharply at the edges of the clusters and stabilizes at a high level in the middle of the respective periods. The boundary points of a cluster may therefore be considered as transitions to neighboring clusters, i.e., to other times. Such assignments allow for the identification of statistically similar households in future work, thereby simplifying the analysis of entire networks.  
   \begin{table}[h!]
\caption{Time Intervals and Clusters}
    \centering
    \begin{tabular}{|c|c|c|c|c|}
        \hline
        0:30-6:00 & 6:30-8:00 & 8:30-15:30 & 16:00-22:00 & 22:30-24:00 \\ \hline
        1 & 2 & 1 & 2 & 1 \\ \hline
    \end{tabular}
    \label{tab3}
\end{table}

Next, we analyze the dependency among subsequent observations in the same cluster.
This means we will examine the periods 0:30-6:00, 6:30-8:00, 8:30-15:30, 16:00-22:00, and 22:30-24:00 separately. We model the dependency structure for each of these periods separately using Vine Copulas. The corresponding tree structure arises naturally, since we connect neighboring time points. The analysis of the dependencies using the R package \textit{VineCopula} yields various bivariate copula models.
 \textit{VineCopula} chooses the best copula between two variables based on the Akaike information criterion.


The chosen copula models are  \textit{(rotated) Joe, (rotated) BB8, t, (rotated) Gumbel, Independence, (rotated) Clayton, Frank, (rotated) Tawn type 1, (rotated) Tawn type 2, Gaussian}, i.e., we mainly observe tail dependencies here, which means that high or low values in active power at different times often are related.\footnote{Due to space limitations of this note, we cannot detail the properties of these copula models here, we refer to \cite{Copula} for an introduction.} 

\begin{figure}[t] 
\centering 
\includegraphics[width=0.49\textwidth,height = 3.5cm]{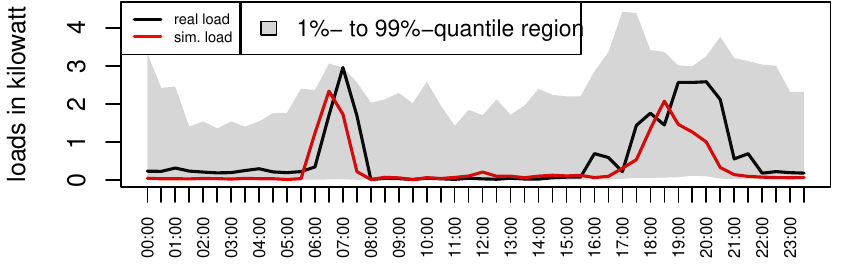} 
\caption{\label{Fig:l1}Simulated and real load profiles with quantiles plotted over time} 
\end{figure}

Using inverse transform sampling we now derive the cumulative distribution function (c.d.f.) corresponding to the kernel density estimate for each period.
Combining these univariate  c.d.f.s using the associated vines, we can now generate individual load profiles for this household. By allowing only values within a prespecified quantile range of the marginal distributions, we can focus on statistically extreme values in certain periods of the day. A simulated load profile is shown in Figure \ref{Fig:l1}. The shaded region represents the area between the 1\%-quantile and the 99\%-quantile. The thin line represents real load data for  comparison.

\begin{figure}[t] 
\centering 
\includegraphics[width=0.49\textwidth,height = 3.5cm]{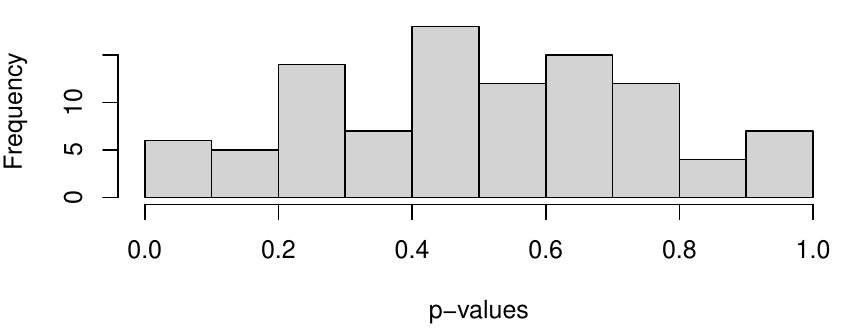} 
\caption{\label{Fig: p}Histogram of p values for household \#1} 
\end{figure}

We also generated load profiles for other households in the dataset with the same procedure as for household \#13. Like for household \#13 the KDE for the other households is based on 156 observations. To compare our simulated load profiles with the real load profiles, we apply a permutation test with 10000 repetitions. The statistical features considered are the $25\%, 50\%, 75\%, 95\%$ percentiles and the maximum of the real and simulated data, respectively. We applied the test 100 times and obtained an average p-value of 0.4 with a standard deviation of 0.33. The distribution of the p-values obtained thus resembles a uniform distribution over $[0,1]$, which indicates that our simulation procedure generates load profiles quite similar to the real ones. A histogram of the observed p values is given in Figure \ref{Fig: p}. One can observe in this data example that the proposed combination of kernel density estimation and vine copulas provides a promising avenue to probabilisitic time-series forecasting. 

\section{Conclusions and Outlook} 

This paper has taken first steps towards combining  kernel density estimation with copulas toward load prediction in distribution grids. Specifically, we fitted time-varying probability distributions to describe electricity consumption on a 30 minutes time grids during certain periods of the day.  Our approach captures temporary peak loads, as opposed to standard load profiles.
We have identified different time periods with similar consumption characteristics using clustering based on Hellinger distances between density estimates.  Future work includes the use of different distance measures, such as the Wasserstein distance and alternative Vine Copula structures, i.e., 48-dimensional Vines.

To obtain a realistic time course of energy consumption, we have modeled temporal dependencies within periods with similar characteristics using vine copulas. An open question is the transition between these states, which have been considered independently here. A different modeling of these transitions could lead to even more realistic load profiles. Moreover, an in-depth comparison to existing quantile based approaches \cite{QuantileReg3} is subject of future research. 

We note that the employed data-driven statistical methods transfer readily to the estimation of aggregated loads at substations or at grid interconnections. Yet, currently there appears to be a lack of suitable open-source datasets.


\begin{thebibliography}{00}
\bibitem{forecasting} J.A. González Ordiano, S. Waczowicz, V. Hagenmeyer and R. Mikut, 2018, "Energy forecasting tools and services", WIREs Data Mining Knowl Discov, 8: e1235.
\bibitem{prohabilistic} T. Hong, P. Pinson, S. Fan, H. Zareipour, A. Troccoli, R. J. Hyndman, "Probabilistic energy forecasting: Global Energy Forecasting Competition 2014 and beyond", Int J Forecasting, Volume 32, Issue 3, 2016, pp. 896-913.
\bibitem{kernelSVR} J. Che, J. Wang, "Short-term load forecasting using a kernel-based support vector regression combination model", Applied Energy, Volume 132, 2014, pp. 602-609.
\bibitem{kernelNonPara} B. Gu, T. Zhang, H. Meng, J. Zhang, "Short-term forecasting and uncertainty analysis of wind power based on long short-term memory, cloud model and non-parametric kernel density estimation", Renewable Energy, Volume 164, 2021, pp. 687-7087.
\bibitem{SplineQ} K. Wang, Y. Zhang, F. Lin, J. Wang and M. Zhu, "Nonparametric Probabilistic Forecasting for Wind Power Generation Using Quadratic Spline Quantile Function and Autoregressive Recurrent Neural Network", in IEEE Trans  Sus En, vol. 13, no. 4, pp. 1930-1943, Oct. 2022.
\bibitem{SplineSVR} Y. He, H. Li, S. Wang, X. Yao, "Uncertainty analysis of wind power probability density forecasting based on cubic spline interpolation and support vector quantile regression", Neurocomputing, Volume 430, 2021, pp. 121-137.
\bibitem{QuantileReg} A. Faustine and L. Pereira, "FPSeq2Q: Fully Parameterized Sequence to Quantile Regression for Net-Load Forecasting With Uncertainty Estimates," in IEEE Trans Smart Grid, vol. 13, no. 3, pp. 2440-2451, May 2022.
\bibitem{QuantileReg2} V. Jensen, F. M. Bianchi and S. N. Anfinsen, "Ensemble Conformalized Quantile Regression for Probabilistic Time Series Forecasting," in IEEE Trans  Neur Net  Learn Sys, vol. 35, no. 7, pp. 9014-9025, July 2024. 
\bibitem{QuantileReg3} J. A. González Ordiano, L. Gröll, R. Mikut, V. Hagenmeyer, "Probabilistic energy forecasting using the nearest neighbors quantile filter and quantile regression", Int J Forecasting, Volume 36, Issue 2, 2020, pp. 310-323.
\bibitem{PVNN} J.A. González Ordiano, S. Waczowicz, M. Reischl, R. Mikut, V. Hagenmeyer, "Photovoltaic power forecasting using simple data-driven models without weather data", Comput Sci Res Dev 32, 2017, pp. 237–246.
\bibitem{Copula1} R. J. Bessa, V. Miranda, A. Botterud, Z. Zhou, J. Wang, "Time-adaptive quantile-copula for wind power probabilistic forecasting", Renewable Energy, Volume 40, Issue 1, 2012, pp. 29-39.
\bibitem{Copula2} Z. Wang, W. Wang, C. Liu, Z. Wang and Y. Hou, "Probabilistic Forecast for Multiple Wind Farms Based on Regular Vine Copulas," in IEEE Trans  Pow Sys, vol. 33, no. 1, pp. 578-589, Jan. 2018.
\bibitem{SLP} D. Peters, R. Völker, F. Schuldt and K. von Maydell, "Are standard load profiles suitable for modern electricity grid models?," 2020 17th International Conference on the European Energy Market (EEM), Stockholm, Sweden, 2020, pp. 1-6.
\bibitem{SLP2} R. Bitterer and B. Schieferdecker, "Repräsentative VDEWLastprofile AktionsplanWettbewerb, M-32/99," VDEW, Frankfurt, Germany, Tech. Rep. M32-99, 2001.
\bibitem{Generator} N. Pflugradt, J. Teuscher, B. Platzer and W. Schufft, "Analysing low-voltage grids using a behaviour based load profile generator", REPQJ, Vol.1, No.11, pp. 361-365, March 2013.
\bibitem{Generator2} B. V. M. Vasudevarao, M. Stifter and P. Zehetbauer, "Methodology for creating composite standard load profiles based on real load profile analysis," 2016 IEEE PES Innovative Smart Grid Technologies Conference Europe (ISGT-Europe), Ljubljana, Slovenia, 2016, pp. 1-6.
\bibitem{SLPProblem} M. Stokes, "Removing barriers to embedded generation: A fine-grained
load model to support low voltage network performance analysis," Ph.D. dissertation, Inst. Energy Sustain. Develop., De Montfort Univ., Leicester, U.K., 1993.
\bibitem{b1} E. L. Ratnam, S. R. Weller, C. M. Kellett and A. T. Murray, ``Residential load and rooftop PV generation: an australian distribution network dataset,'' Int J  Sus En 8, vol. 36, pp. 787--806, 2017.
\bibitem{b2} S. Weglarczyk, Kernel density estimation and its application, ITM Web of Conferences, vol. 23, 2018.
\bibitem{Vaart} A. W. van der Vaart, "Asymptotic Statistics", Cambridge Series in Statistical and Probabilistic Mathematics, Cambridge University Press, 1998.
\bibitem{Sil} P. J. Rousseeuw, "Silhouettes: A graphical aid to the interpretation and validation of cluster analysis", Journal of Computational and Applied Mathematics, Nr. 20, pp. 53–65, 1987.
\bibitem{b3} C. Czado and T. Nagler, ``Vine copula based modeling,'' Ann Rev Stat Appl, vol. 9, pp. 1--27, 2021.
\bibitem{b4} A. Sklar, ``Fonctions de r´epartition `a n dimensions et leurs marges,'' Publications de l’Institut de Statistique de L’Universit´e de Paris, vol. 8, pp. 229--231, 1959.
\bibitem{b5} M. C. Jones, J. S. Marron and S. J. Sheather, ``A brief survey of bandwidth selection for density estimation,'' J   Amer Stat Asc, Vol. 91, No. 433, pp. 401--
407, 1996.
\bibitem{b6} H. Akaike, ``A new look at the statistical model identification,'' IEEE Trans  Aut Con, vol. 19 (6), pp. 716--723, 1974
\bibitem{Copula} P. Jaworski, F. Durante, W. K. Härdle and T. Rychlik, "Copula Theory and Its Applications", 1st ed., Springer Berlin, Heidelberg, July 2010.
\end{thebibliography}
\end{document}